\documentclass[pdflatex,sn-mathphys-num]{sn-jnl}

\usepackage{graphicx}%
\usepackage{multirow}%
\usepackage{amsmath,amssymb,amsfonts}%
\usepackage{amsthm}%
\usepackage{mathrsfs}%
\usepackage[title]{appendix}%
\usepackage{xcolor}%
\usepackage{textcomp}%
\usepackage{manyfoot}%
\usepackage{booktabs}%
\usepackage{algorithm}%
\usepackage{algorithmicx}%
\usepackage{algpseudocode}%
\usepackage{listings}%
\usepackage{caption}
\usepackage{subcaption}
\usepackage{enumitem} 

\theoremstyle{thmstyleone}%
\theoremstyle{thmstyletwo}%

\theoremstyle{thmstylethree}%

\begin{document}

\title[Article Title]{Geodesic completeness of singularity-free gravity}


\author*[1]{\fnm{Vasil} \sur{Todorinov}}\email{v.todorinov@uleth.ca}



\affil*[1]{\orgdiv{Physics \& Astronomy Department}, \orgname{University of Lethbridge}, \orgaddress{\street{4401 University Drive}, \city{Lethbridge}, \postcode{T1K 3M4}, \state{Alberta}, \country{Canada}}}




\abstract{Generalizing the gravitational potential first proposed in \cite{Culetu:2014lca}, we derive a large class of relativistic singularity-free theories of gravity, which reduce to flat spacetime at large distances. We verify that for the chosen gravitational potential the force and the spacetime curvature resolve the singularity and vanish at large distances. We show that those singularity-free black-hole solutions generically have a two-horizon structure. Furthermore, we show that there is a subclass of potentials that produce a geometries that are geodesically complete through the origin. We discuss the implications of the effects resulting from such theories and show that black-hole solutions are predicted to have a minimum allowed mass.}




\maketitle

\section{\label{sec:Int}Introduction}

Classical general relativity (GR) is extraordinarily successful on macroscopic scales, yet it is also well known to have curvature singularities.  In this sense,    singularities are not merely ``pathological solutions'', but rather signal a breakdown of the classical description of gravity and motivate the exploration of modified theories of gravity \cite{Penrose1965PRL,Hawking1966RSPA1,Geroch:1966ur,Ellis:1968vy,HawkingPenrose1970RSPA,Geroch1968AnnPhys,Ellis:1977pj,Tipler:1977eb,Kriele:1990xy,Senovilla1998GRG,Senovilla:2014gza,Borde:1993xh,Borde:1994ai,Borde:2001nh,Ferreira:2025rje,Borde:1996pt,Hawking1976PRD,Hawking:1982dj,Hartle:1983ai,Vilenkin:1982de,Mukhanov:1991zn,Carlip:2001wq,Kuntz:2019lzq,Gielen:2016fdb,HorowitzSteif1990PRL,Horowitz:1990ap,Bojowald:2001xe,Ashtekar:2006rx,Bojowald:2005qw,Bojowald2008LRR,Ashtekar:2011ni,Ashtekar:2005qt,Modesto:2004xx,Gambini:2008dy,Stelle:1976gc,Starobinsky:1980te,Deser:1998rj,Banados:2010ix,BeltranJimenez:2017doy,Capozziello:2011et,Nojiri:2010wj,Nojiri:2008fk,Frolov:2008uf,Appleby:2009uf,Biswas:2005qr,Biswas:2010zk,Biswas:2011ar,Nash:2023zza,Cotton:2021tfl}. 

However, \emph{regularity of curvature invariants alone is not enough}.  Recent analyses have emphasized that several popular regularized black-hole metrics can remain \emph{geodesically incomplete}. Consistent analytic extensions and non-analyticity near the origin can introduce a genuine ambiguity in the procedure \cite{Zhou:2022yio}.  Since geodesic completeness is the operational criterion for the absence of classical singularities, any proposal that regularizes curvature but leaves the maximal extension ambiguous risks shifting the singularity problem rather than resolving it.  This observation provides a concrete target for model-building, with the caveat that any modification to the Newtonian potential should be sufficiently controlled so that causal structure and geodesic evolution can be analyzed.

 Additionally, there is renewed interest in constructing singularity-free frameworks that modify gravity already at the level of an effective potential. For example, in a recent work  \cite{Das:2024tme} the authors propose a classical theory designed to be singularity-free at short distances and to reduce to GR at large distances. Within that framework, we are naturally led to exponentially-suppressed potentials that weaken gravity close to the origin.  Independently, minimum-length ideas from quantum-gravity phenomenology, notably generalized uncertainty principles (GUP) \cite{Adler1999-db,Adler_2001,
	Ali_2014,
	Ali_2015,
	Alonso_Serrano_2018,Amati1989-gs,
	AMELINO_CAMELIA_2002,
	Bargue_o_2015,Bambi2007-te,Bawaj_2015,
	Bojowald2011-bb,Bolen2005-jq,
	Bosso2018,
	Bosso:2018uus,
	Bosso:2019ljf,
	Burger_2018,
	Capozziello:1999wx,
	Chang:2011jj,
	Cortes:2004qn,
	Costa_Filho2016-ox,
	Dabrowski2019-cb,
	Dabrowski2020-kk,
	Das2008,
	Das:2010zf,
	DAS2011596,
	Das2014,
	Das_2019,
	Das_2020,
	Garcia-Chung:2020zyq,
	Giddings2020-xz,
	Hamil2019-qh,
	Hossenfelder:2012jw,
	Kempf1995-ka,
	Kober:2010sj,KONISHI1990276,
	MAGGIORE199365,
	Maggiore:1994,
	Marin:2013pga,
	Mead1966-xj,
	Moradpour2021-jy,
	Mureika2019-lf,
	Myung_2007,
	Park2008-uj,
	Scardigli1999-ne,
	Snyder:1946qz,
	Sprenger_2011}, also motivate non-local or exponential softening in effective interactions.  The relativistic generalized uncertainty principle (RGUP) provides a Lorentz-covariant implementation of a minimum length and supplies an explicit, internally consistent route from microscopic deformations to effective dynamics \cite{Todorinov2018-xi,Nenmeli:2021orl}.  While these approaches are not identical, they agree that short-distance physics should \emph{regularize} the gravitational potential in a way that is controllable, ideally analytic, and testable via robust geometric diagnostics.
    
This works is structured as follows. In Section \ref{Sec:PotentialFamily} we  study an analytically tractable, exponentially-regulated family of potentials parameterized by a positive exponent $\ell$ and a short-distance length scale $\lambda$. We treat the potential as an effective phenomenological proposal and focus on \emph{invariants of causal structure} that are insensitive to coordinate choices. In Section \ref{sec:horizons} we determine the corresponding horizon structure and its parametric dependence and compute a minimum black-hole mass resulting from the added small length scale $\lambda$. In Section \ref{sec:GeodesicStructure} we study the geodesic completeness of the proposed theory. Following \cite{Zhou:2022yio,Das:2024tme} we compute radial null geodesics and the associated tortoise coordinates, and  construct Eddington--Finkelstein causal diagrams to diagnose whether the proposed short-distance behaviour leads to a geodesically complete maximal extension. Furthermore, we address the question of whether curvature is finite and the resulting spacetime admits an unambiguous extension, in addition to being a physically acceptable geodesic evolution. 

Consequently, whenever we adopt a specific identification (e.g.\ $g_{tt}=-(1+2V/c^2)$ in a weak-field motivated closure), we treat it as an explicit modelling choice and test the resulting causal predictions against the specified criteria.
This  separation between the phenomenological potential $V(r)$ and the geometric/causal diagnostics is, in our view, essential if exponentially-softened ``regular black hole'' proposals are to be assessed critically and compared meaningfully across the literature. Although the construction is initially motivated by a Newtonian-like potential, the analysis is not restricted to Newtonian physics. The embedding of the modified potential into a metric ansatz and the resulting framework is fully spacetime-based and admits a covariant geometric interpretation through horizon structure, null geodesics, and causal extensions. In Section \ref{sec:fR_reconstruction_rho} we reconstruct a $f(R)$ gravitational action which reproduces the horizon structure discussed in the previous Section. The results and future directions of this work are discussed in Section \ref{sec:Conclusion}. 

\section{Expanded Exponentially Regularized Potential}\label{Sec:PotentialFamily}

The potential given in \cite{Culetu:2014lca} is comprised of two terms a de-Sitter core $R_S/2r$ which provides the asymptotic flatness and reproduces the behaviour of the universe at cosmological scales,  and the second term is an exponential "regulator" term $e^{-\frac{\lambda^2}{2 R_S r}}$ which possibly resolves the singularity at the origin by making gravity asymptotically free. 
\begin{equation}\label{eq:pureCuletu}
	V_{C_0}(r)=-\frac{R_S}{2r}e^{-\frac{\lambda^2}{2 R_S r}}\,,
\end{equation}
where $R_S= 2\,G\,M/c^2$ is the Schwarzschild radius and $\lambda$ is a constant with dimensions of length. The potential in Eq. \eqref{eq:pureCuletu} is a single representative of a whole family of potentials parametrized by the powers $q,p$ of the distance from the source. In order to explore the family of potentials of which $V_C$ is a particular case we write the general member of the family as
\begin{equation}\label{eq:generalanzatz}
	V_C(\rho,\eta,p,q)=-\frac{1}{2}\left(\eta\rho\right)^p e^{-\frac{\eta}{2}\,\rho^q}\,,
\end{equation}
where we defined  $\eta \equiv\lambda/R_S$ at the ratio between the two lengths scales in our theory, and $\rho=r/\lambda$ is a dimensionless radial coordinate.  The original potential proposed in \cite{Culetu:2014lca}  is recovered by taking the $p=q=-1$ values for the powers. The generalized form of the   potential proposed in Eq. \eqref{eq:generalanzatz} is chosen to reproduce the asymptotic behaviour of the Schwarzschild metric at large distances, while resolving the singularity at the origin. Additionally, the power law in the exponential term in  Eq. \eqref{eq:generalanzatz} allows for the reconstruction of a spacetime that is geodesically complete rather than shifting the singularity. 

Exploring  the limits of the members of the potential family in Eq. \eqref{eq:generalanzatz}
\begin{equation}
	\lim_{\rho\rightarrow 0^+}V_C(\rho)\,,\quad\text{and}\quad \lim_{\rho\rightarrow\infty}V_C(\rho)\,.
\end{equation}
allows us to narrow down the range of powers $(q,p)$ for which the potential is zero at infinity and finite at zero.  As seen from Table \ref{tab:Assymtotics} the admissible possibilities are that the powers $p$ and $q$ are either both strictly positive or both strictly negative. We will refer to the strictly negative case as $\tilde V_C$.

The Taylor expansion of the extended potential with $p<0$ and $q<0$ at infinity is:
\begin{align}
	\tilde V_C(r)&= -\frac{1}{2}\left(\eta \rho\right)^p e^{-\frac{\eta}{ 2 }\left(\rho\right)^q}\\&= -\frac{1}{2}\left(\eta \rho\right)^p \sum_{n=0}^{\infty}\frac{\left(-1\right)^n\left(\frac{\eta}{ 2 }\rho^q\right)^n}{n!}= \sum_{n=0}^{\infty}\frac{\left(-1\right)^{n+1}\eta^{n+p}\rho^{q\,n+p}}{2^{n+1}\,n!}\label{CuletuTaylor}\,,
\end{align}
while for the case of $p>0$ and $q>0$ the Taylor series at infinity does not exist.
For this reason the branch of the potential family we choose to explore further is the one where $p$ and $q$ are strictly negative. 
\begin{table}[h!]
	\centering
	
	\begin{tabular}{@{}ccccc@{}}
		\toprule
		& & $ \lim_{r\rightarrow 0^+}V_C(r)$ & $\lim_{r\rightarrow\infty}V_C(r)$ & $V(r)$ at $r\rightarrow \infty$\\
		\hline
		$p<0$  & $q<0$ & $0$&$0$ & $V_C(r)\propto \rho^p$\\
		$p<0$  & $q\geq 0$ & $-\infty$&$0$  & $q=0\rightarrow V_C(r)\propto \rho^p$\\
		$p=0$  & $q>0$ & $-1/2$&$0$  & N/A\\
		$p=0$  & $q<0$ & $0$&$-1/2$  & $V_C(r)\propto \rho^{q}$\\
		$p>0$  & $q\leq 0$ & $0$ & $-\infty$ & $V_C(r)\propto \rho^{p}$\\
		$p>0$  & $q>0$ & $0$ & $0$  & N/A\\
		\botrule
	\end{tabular}
	\caption{Asymptotic behaviour of the different branches of the potential family shown in Eq. \eqref{eq:generalanzatz} categorized by the powers $p$ and $q$ in the potential.}
	\label{tab:Assymtotics}
\end{table}

As Eq. \eqref{CuletuTaylor} shows the leading order of the Taylor expansion at infinity is proportional to $\rho^p$. Thus, in order to have a $1/\rho$ term at infinity we need the power $p=-1$. Furthermore, we will rewrite the potential replacing $\rho^q$, for $q<0$, by $1/\rho^\ell$ with $\ell>0$ which yields the general expression for the potential
\begin{equation}\label{Eq:GeneralPotnetial}
	V_C(\rho)= -\frac{1}{2\eta\rho} e^{-\frac{\eta}{ 2 \rho^\ell }},\qquad \ell>0\,,\, \eta>0
\end{equation}
The equation above shows the only branch of the Potential family presented in Eq. \eqref{eq:generalanzatz} which reproduces the long-distance asymptotic structure of the Schwarzschild spacetime, while resolving the singularity at the origin. This provides a great candidate for reconstructing the action of a asymptotically and singularity-free theory of General Relativity.

\section{Horizons of the generalized potential}\label{sec:horizons}

It is worth stating again that although the form of the potential chosen is constructed in the non-covariant framework, we can reconstruct the covariant theory by inserting the resulting potential in a weak field limit and searching for an action reproducing that behaviour. This leads to a fully covariant theory of asymptotically free gravity. We need to make sure that the resulting spacetime not only satisfies the asymptotic conditions we imposed but is geodesically complete through the centre of the source, thereby ensuring that the singularity has not merely been pushed into the $\rho<0$ domain.    
\subsection{Horizon Structure}
Considering the potential form given in Eq. \eqref{Eq:GeneralPotnetial} we adopt the metric ansatz 
\begin{equation}\label{eq:g00dimless}
    g_{tt}=-c^2 f(\rho), \quad f(\rho)\equiv 1+2V_C(\rho)=1-\frac{1}{\eta\rho}\exp\left(-\frac{\eta}{2\rho^\ell}\right)\,.
\end{equation}
This form is motivated by the usual weak-field identification $g_{tt}\simeq -c^2\left(1+2\Phi/c^2\right)$, where in our case $V_C=\Phi/c^2$. Horizons occur at $g_{tt}=0$, i.e.
\begin{equation}
	\frac{1}{\eta\rho}\exp\!\left[-\,\frac{\eta}{2\rho^\ell}\right]=1.
	\label{eq:h-eq}
\end{equation}
Let $y\equiv \rho^\ell$ and $z\equiv -\dfrac{\eta}{2y}$ so that $y=-\dfrac{\eta}{2z}$.
From Eq.\eqref{eq:h-eq} we find
\begin{equation}
	(\ell z)\,e^{\ell z}=-\frac{\ell}{2}\,\eta^{\,\ell+1}.
\end{equation}
Using the Lambert $W$ function defined by $W e^{W}=x$ \cite{Corless:1996zz},
\begin{equation}
	\ell z=W_k\!\!\left(-\frac{\ell}{2}\eta^{\ell+1}\right),\qquad k\in\{0,-1\}.
\end{equation}
Since $z=-\eta/(2\rho^\ell)$, we obtain the horizon radii
\begin{equation}
	\rho_h^{(k)}(\eta,\ell)
	=\eta^{-1}\left[\frac{\alpha}{-\,W_k(-\alpha)}\right]^{\!1/\ell},
	\label{eq:rhoh-sol}
\end{equation}
where $\alpha\equiv \frac{\ell}{2}\,\eta^{\,\ell+1}$ and  $k\in\{0,-1\}$ parametrizes the two branches of the Lambert function.

Real horizons correspond to real values of $W_k$. The real branches satisfy $W_0(x)\in[-1,\infty)$ and $W_{-1}(x)\in(-\infty,-1]$ for $x\in[-e^{-1},0)$ \cite{Corless:1996zz}.
Thus we must have
\begin{equation}
	-\alpha\in[-e^{-1},0) \quad \Longleftrightarrow\quad\alpha=\frac{\ell}{2}\eta^{\,\ell+1}\ \le\ \frac{1}{e}\ .
	\label{eq:existence}
\end{equation}
Therefore
\begin{equation}
	\label{eq:mass-threshold1}
	\eta \;\le\; \left(\frac{e\,\ell}{2}\right)^{-1/(1+\ell)}\,,
\end{equation}
Equivalently
\begin{equation}
	\label{eq:mass-threshold}
	R_S \;\ge\; \lambda\,\Big(\tfrac{e\,\ell}{2}\Big)^{\!1/(1+\ell)}\,.
\end{equation}

The multiplicity of the horizons is determined by the bound in Eq. \eqref{eq:mass-threshold}.
In the case where the bound in Eq. \eqref{eq:mass-threshold} holds strictly, there are \emph{two} real solutions:
an outer horizon from $W_0$ and an inner horizon from $W_{-1}$. For macroscopic $R_S\gg \lambda$, we have $\alpha\ll1$, so both real branches are available.
A special case of the above condition is when the equality holds, then the two horizons collapse into single degenerate or extremal, horizon. Lastly, if the bound fails, no real solution exists and the horizons disappear and we end up with a horizon-less solution.

\subsection{Minimum Schwarzschild radius and Black-Hole mass}
An extremal or in other words degenerate horizon occurs at $f(\rho_\star)=0=f'(\rho_\star)$.
Differentiating $f$ yields
\begin{equation}
	f'(\rho)=\frac{e^{-\eta/(2\rho^\ell)}}{\eta}\,\rho^{-\ell-2}
	\left(\rho^\ell-\frac{\ell\,\eta}{2}\right).
	\label{eq:fp}
\end{equation}
At a horizon $\rho=\rho_h$, Eq. \eqref{eq:h-eq} gives $e^{-\eta/(2\rho_h^\ell)}=\eta\rho_h$.
Thus $f'(\rho_h)=0$ is equivalent to
\begin{equation}
	\rho_h^\ell=\frac{\ell\,\eta}{2}.
	\label{eq:ext-cond1}
\end{equation}
Substituting Eq. \eqref{eq:ext-cond1} back into Eq. \eqref{eq:h-eq} gives
\begin{equation}
	e^{-\eta/(2\rho_h^\ell)}=e^{-1/\ell}=\eta\,\rho_h,
\end{equation}
so $\rho_h=e^{-1/\ell}/\eta$.
Raising $\rho_h$ to the power $\ell$ and using Eq. \eqref{eq:ext-cond1} we get
\begin{equation}
	\frac{e^{-1}}{\eta^\ell}=\frac{\ell\,\eta}{2}
	\Rightarrow \alpha=\frac{\ell}{2}\,\eta^{\ell+1}=\frac{1}{e}\,.
\end{equation}
Therefore, the unique double horizon occurs at
\begin{equation}
	\rho_\star=\frac{e^{-1/\ell}}{\eta},\qquad
	\rho_\star^\ell=\frac{\ell\,\eta}{2}.
	\label{eq:rhostar}
\end{equation}

\begin{figure}
	\centering
	\includegraphics[width=\textwidth]{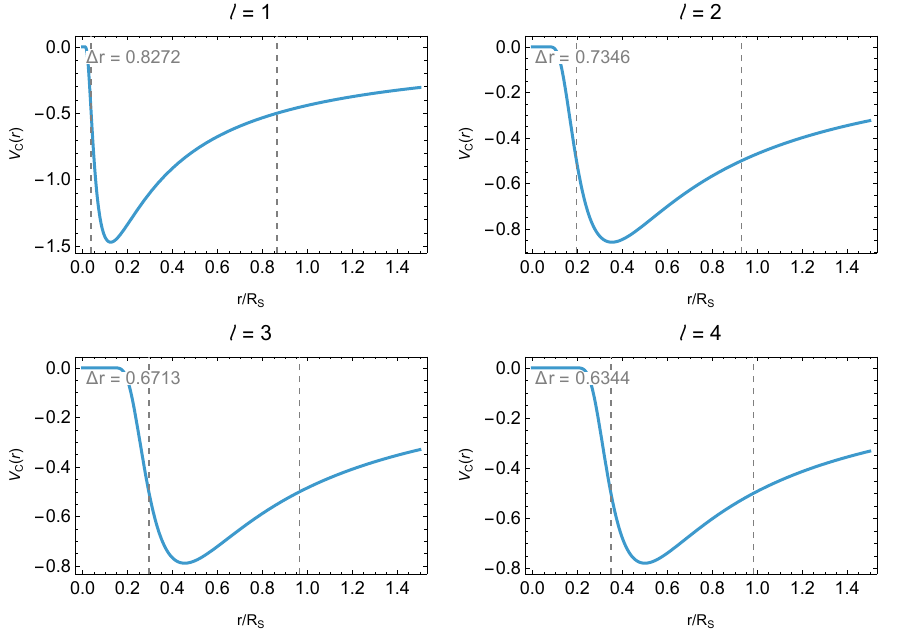}
	\caption{The blue lines in the figure show the gravitational potential and the vertical dashed lines show the horizon structure for different cases of the powers $\ell$ in the asymptotically free potential in Eq. \eqref{Eq:GeneralPotnetial}. The distance between the two horizons is denoted in the corner. The desired features of the potential are clearly visible.}
	\label{fig:Horizons}
\end{figure}

To relate $\eta$ to mass, set $\eta=\lambda/R_S$ with
$R_S=2GM/c^2$, by construction $\eta$ is
dimensionless. The extremality condition $\alpha=1/e$ gives
\begin{equation}
	\eta_{\max}=\left(\frac{2}{e\,\ell}\right)^{\!\frac{1}{\ell+1}},
\end{equation}
\begin{equation}
	R_{S,\min}=\frac{\lambda}{\eta_{\max}}
	=\lambda\left(\frac{e\,\ell}{2}\right)^{\!\frac{1}{\ell+1}}.
	\label{eq:RSmin}
\end{equation}
Hence the \emph{minimum mass} allowed by the two-horizon structure is
\begin{equation}
	M_{\min}(\ell,\lambda)=\frac{c^2}{2G}\,R_{S,\min}
	=\frac{c^2\lambda}{2G}\left(\frac{e\,\ell}{2}\right)^{\!\frac{1}{\ell+1}}\,.
	\label{eq:Mmin}
\end{equation}
In order to provide some intuitive understanding of the relationship between the additional length scale and the minimum mass we can set $\lambda=L_{\text{Planck}}$, in which case 
\begin{equation}
	M_{\min,\lambda}(\ell)
	=\frac{m_P}{2}\left(\frac{e\,\ell}{2}\right)^{\!\frac{1}{\ell+1}},
\end{equation}
where $m_P=\sqrt{\hbar c/G}$ is the Planck mass ($R_S(m_P)=2L_{\text{Planck}}$).

In the extremal limit $M\rightarrow M^{+}_{\min}$, the two horizons coalesce and the surface gravity vanishes. As $\ell\to\infty$, $M_{\min}(\ell)\to \tfrac{1}{2}m_P$.
For representative values: $\ell=1\Rightarrow M_{\min}\approx 0.58\,m_P$;
$\ell=2\Rightarrow M_{\min}\approx 0.70\,m_P$.
The two-horizon structure enforces a sharp threshold for the existence of the horizons i.e.
$\alpha\le 1/e$.
Equivalently, there is a \emph{minimum mass} $M_{\min}(\ell)$ given by Eq. 
\eqref{eq:Mmin}. Below that minimum mass $M<M_{\min}$ the time component of the metric $g_{00}$ is negative everywhere, equivalently the lapse function satisfies $f(r)>0$ everywhere, thus
no horizon forms.
	\begin{figure}[htbp]
		\centering
		\begin{subfigure}[b]{0.51\textwidth}
			\includegraphics[width=\textwidth]{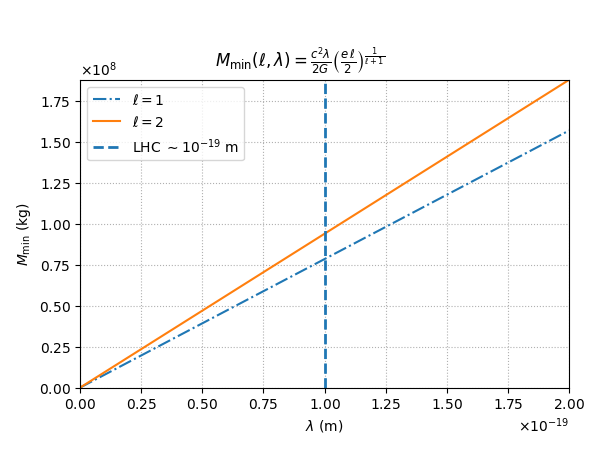}
			\caption{Near the LHC scale}
			\label{fig:sub1}
		\end{subfigure}
		\hfill
		\begin{subfigure}[b]{0.45\textwidth}
			\includegraphics[width=\textwidth]{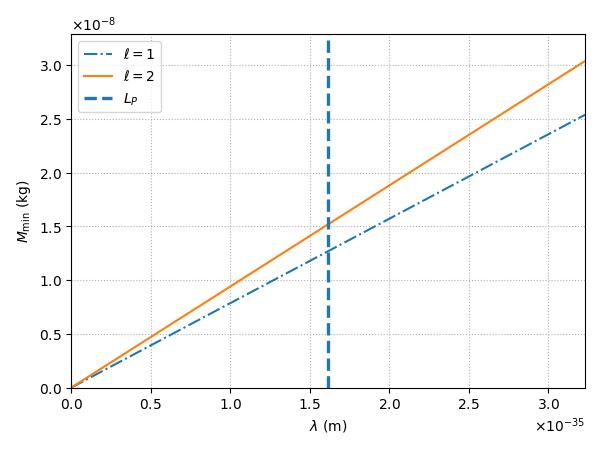}
			\caption{Near the Planck scale}
			\label{fig:sub2}
		\end{subfigure}
	\caption{The solid and dash-dotted lines show the value of the minimum black-hole mass as a function of the additional length scale $\lambda$ introduced in Eq.\eqref{Eq:GeneralPotnetial}.In panel \ref{fig:sub1}, the vertical dashed line marks $\lambda$ equal to the LHC scale, while in panel \ref{fig:sub2} the dashed line marks $\lambda$ equal to the Planck length.}
	\label{fig:MinMass}
\end{figure}
\subsection{Asymptotics for Small $\alpha$ }

Near extremality $\left(\alpha\to 1/e^-\right)$, the two horizons coalesce with a square-root splitting controlled by the standard expansion of $W$ about its branch point at $-1/e$ \cite{Corless:1996zz}.

When $\alpha\ll 1$ or in other words $\eta\ll 1$, the principal branch admits the expansion
\begin{equation}
	W_0(-\alpha)=-\alpha-\alpha^2+O(\alpha^3)\,,
\end{equation}
while $W_{-1}(-\alpha)\sim \ln\alpha$ ($\to -\infty$).

Using Eq. \eqref{eq:rhoh-sol} we get the following expressions for the locations of the two horizons
\begin{equation}\label{eq:rhopm}
	\rho_+ = \eta^{-1}\!\left[1-\frac{\alpha}{\ell}+O(\alpha^2)\right]\qquad
	\rho_- = \eta^{-1}\!\left(\frac{\alpha}{|\ln\alpha|}\right)^{\!1/\ell}\!\ll \rho_+,
\end{equation} 
or substituting the definitions $\eta\equiv \lambda/R_S$ and  $\rho\equiv r/\lambda$  we restore the physical distance to the horizons
\begin{align}
	r_+ &= R_S\left(\frac{\alpha}{-\!W_0(-\alpha)}\right)^{\!1/\ell}
	= R_S\left(\frac{1}{1+\alpha+O(\alpha^2)}\right)^{\!1/\ell}
	= R_S\left[1-\frac{\alpha}{\ell}+O(\alpha^2)\right],\label{eq:rplus-asymp}\\
	r_- &= R_S\left(\frac{\alpha}{-\!W_{-1}(-\alpha)}\right)^{\!1/\ell}
	\approx R_S\left(\frac{\alpha}{|\ln\alpha|}\right)^{\!1/\ell}
	\ll R_S.\label{eq:rminus-asymp}
\end{align}
Eq. \eqref{eq:rplus-asymp} shows the outer horizon lies slightly \emph{inside} $R_S$, with a fractional shift $\sim \alpha/\ell$. In other words the inner horizon is parametrically smaller than the outer one.  This is reassuring because for small $\alpha \rightarrow 0$ which corresponds to small $\eta \rightarrow 0$ in turn $R_S\gg\lambda$ the radius of the outer horizon reduces to the Schwarzschild radius, and the radius of the inner horizon collapses to zero.

\subsubsection{Admissible power by regime} 
In order for us to accurately reproduce a covariant theory of gravity a specific value for the power $\ell$ is needed. Considering the horizon structure we can draw some conclusions for the expected power $\ell$. Two horizons exist if
\begin{equation}
	\alpha(\ell,\eta)\le \frac{1}{e}.
	\label{eq:existence-inequality}
\end{equation}
The threshold $\alpha=1/e$ defines the \emph{merger} (extremal) values of $\ell$ at fixed $\eta$.
Solving
$(\ell/2)\eta^{1+\ell}=1/e
$
for $\ell$ yields
\begin{equation}
	\ell_k(\eta)=\frac{1}{\ln\eta}\,
	W_k\!\!\left(\frac{2\ln\eta}{e\,\eta}\right),\qquad
	k\in\{0,-1\}\,,
	\label{eq:ell-threshold}
\end{equation}
where $W_0$ and $W_{-1}$ are the two real branches on $[-e^{-1},0)$ \cite{Corless:1996zz}.
When the argument leaves this interval, the corresponding branch ceases to be real.

Write $X(\eta)\equiv \dfrac{2\ln\eta}{e\,\eta}$.
There is a distinguished scale $\eta_c\in(0,1)$ determined by
\begin{equation}
	X(\eta_c)=-\frac{1}{e}
	\quad\Longleftrightarrow\quad
	\eta_c=2\,W_0\!\!\left(\tfrac12\right)\approx 0.703467\,.
	\label{eq:eta-c}
\end{equation}
Using Eq. \eqref{eq:ell-threshold} and considering the sign of $\ln\eta$, we find four distinct cases:
\begin{enumerate}[label=(\roman*)]
	\item $\mathbf{0<\eta<\eta_c}$: Macroscopic Schwarzschild radius $R_S$, $\ln\eta<0$, but $X(\eta)<-1/e$.   
	The equation $\alpha=1/e$ has no real solution.  In other words  
	$\alpha(\ell,\eta)<1/e$ for all $\ell>0$ hence there is no upper bound for $\ell$. Two horizons exist for all powers in the exponent.
	\item $\mathbf{\eta=\eta_c}$:  
	The two $W$ branches coalesce and a unique degenerate value $\ell=\ell_\ast$ exist and solves the equation $\alpha=1/e$.
	\item $\mathbf{\eta_c<\eta<1}$: Then $(\ln\eta<0$, and $X(\eta)\in[-1/e,0)$. 
	Both $W_0$ and $W_{-1}$ are real, thus giving two thresholds for $\ell$
	\begin{equation}
		\ell_{-}(\eta)=\frac{W_{0}(X(\eta))}{\ln\eta}\qquad
		\ell_{+}(\eta)=\frac{W_{-1}(X(\eta))}{\ln\eta},
	\end{equation}
	where $0<\ell_{-}<\ell_{+}$. In this case the  horizons exist for
	\begin{equation}
		\ell\in(0,\ell_{-}(\eta)]\ \cup\ [\,\ell_{+}(\eta),\infty)\ ,\ 
		\label{eq:allowed-band}
	\end{equation}
	and are absent for $\ell\in(\ell_{-},\ell_{+})$.
	This ``forbidden band'' is centred near the stationary point
	$\ell^\star=-1/\ln\eta$ of $\alpha(\ell,\eta)$.
	
	\item $\mathbf{\eta=1}$: Then $(R_S=\lambda$, and thus $\ln\eta=0$.  
	We have the equality in Eq.\eqref{eq:existence-inequality} $\alpha=\ell/2$ increases monotonically, so we have a single upper bound
	\begin{equation}
		\ell\le \frac{2}{e}\approx 0.73576
	\end{equation}
	
\end{enumerate}

\section{Geodesic structure}\label{sec:GeodesicStructure}

We have thus far explored the horizon structure and showed that, for non-extremal black-hole solutions, the proposed potential leads to two horizons. If a true resolution of the singularity is sought, the question about the geodesic completeness of the resulting spacetime still stands. In the following, we address this question and find a further constraint on the parameters of the potential in Eq. \eqref{Eq:GeneralPotnetial}. In Appendix \ref{app:radial-sector-center} we have sketched a quick proof that the radial geodesic is enough for us to draw conclusions about the geodesic completeness of the entire spacetime in the non-rotating spherically symmetric case.  

\subsection{Structure of radial timelike geodesics}
The starting point is the Schwarzschild metric (cf.\cite{Zhou:2022yio})
\begin{equation}\label{eq:metric}
	ds^2=-f(\rho)\,c^2dt^2+f(\rho)^{-1}\lambda^2 d\rho^2+\lambda^2 \rho^2 d\Omega^2,
\end{equation}
where  
\begin{equation}
	f(\rho)=1-\frac{1}{\eta\,\rho}\exp\!\left[-\,\frac{\eta}{2\,\rho^{\ell}}\right]
	=1-\frac{2G M(\rho)}{c^2r},
\end{equation}
where $r=\lambda\rho$. In the weak-field limit $g_{tt}\simeq -(1+2V_C)$ our potential
\begin{equation}
	V_C(\rho)= -\frac{1}{2\eta\,\rho}\exp\!\left[-\,\frac{\eta}{2\,\rho^{\ell}}\right],
	\qquad \ell>0,\ \eta>0,
\end{equation}
implies
\begin{equation}
	f(\rho)=1+2V_C(\rho)
	=1-\frac{1}{\eta\rho}\exp\!\left[-\,\frac{\eta}{2\,\rho^{\ell}}\right],
\end{equation}
with the corresponding mass function
\begin{equation}
	M(\rho)=\frac{\lambda c^2}{2G\eta}\exp\!\left[-\,\frac{\eta}{2\,\rho^{\ell}}\right].
\end{equation}

Following the steps outlined in \cite{Zhou:2022yio}, for radial timelike motion in proper time $\tau$, and with conserved total energy
\begin{equation}\label{Eq:EnergyCondition}
	E:=-\frac{u_t}{c^2}=-\frac{g_{tt}}{c^2}\frac{dt}{d\tau}=f(\rho)\frac{dt}{d\tau}.
\end{equation}
Using the zero angular momentum condition $d\Omega=0$, the normalization becomes 
\begin{equation}
	-c^2=-f(\rho)c^2\left(\frac{dt}{d\tau}\right)^2+
	f(\rho)^{-1}\lambda^2\left(\frac{d\rho}{d\tau}\right)^2.
	\label{eq:norm-expanded-rs-rho}
\end{equation}
Substituting  Eq. \eqref{eq:norm-expanded-rs-rho} in Eq. \eqref{Eq:EnergyCondition}, we obtain
\begin{equation}
	-\frac{E^2}{f(\rho)}+
	\frac{\lambda^2}{c^2}\frac{\dot{\rho}^{\,2}}{f(\rho)}=-1\,.
	\label{eq:norm-reduced-rs-rho}\end{equation}
Theretofore, the radial timelike geodesic equation reads
\begin{equation}
	\frac{\lambda^2}{c^2}\dot\rho^{\,2}=E^2-f(\rho)
	= E^2-1+\frac{1}{\eta\rho}\,e^{-\frac{\eta}{2\rho^\ell}},
	\label{eq:radial}
\end{equation}
where $\cdot \equiv d/d\tau$ denotes the time derivative in proper time $\tau$. The admissible region, in which with kinetic terms is positive, is denoted by $\dot\rho^{\,2}>0$.   The proper time from $\rho_i$ to $\rho$ takes the form 
\begin{equation}
	\tau(\rho)=\frac{\lambda}{c}\int_{\rho}^{\rho_i}\frac{d\rho}{\sqrt{\,E^2-1+\dfrac{1}{\eta\rho}\,e^{-\frac{\eta}{2\rho^\ell}}\,}}.
\end{equation}
Because the exponential regulator vanishes at $\rho=0$ i.e.  $\lim_{\rho\rightarrow 0} e^{-\eta/(2\rho^\ell)}=0$, we arrive at $\lim_{\rho\rightarrow 0}\dot\rho^{\,2}= E^2-1$. Thus,
for geodesics with energy $E>1$ the proper time to the centre is finite, triggering the extension test across the origin $\rho=0$ advocated for in \cite{Zhou:2022yio}. 

\subsection{Extension to \texorpdfstring{$\rho<0$}{?<0}}
The criterion for geodesic completeness is that for geodesics that reach $\rho=0$ in finite proper time, we must extend to negative radius and verify that the geodesic equations remain well-defined.  Otherwise the spacetime is geodesically incomplete \cite{Zhou:2022yio,Hayward:2005gi}. In the following we apply this condition for three different cases for the power $\ell$, which in turn determines the shape of the exponential regulator. 

\subsubsection{Non-integer power}

For radial timelike geodesics, the affine parameter $\,\tau\,$ equals proper time, with conserved specific energy $E$
and no angular momentum $L=0$, the normalization $g_{\mu\nu}\dot x^\mu \dot x^\nu=-1$ gives the result presented in Eq. \eqref{eq:norm-reduced-rs-rho} and the radial geodesic equation in Eq. \eqref{eq:radial}
As $\rho\to 0^+$, the exponential decays faster than any power, hence $\dot\rho^{\,2}\to E^2-1$. If $E>1$, then for some $\rho_0>0$,
\begin{equation}\label{eq:finite-time}
	\int_{0}^{\rho_0}\frac{d\rho}{\sqrt{E^2-1+\dfrac{1}{\eta\rho}e^{-\eta/(2\rho^\ell)}}}
	\;\le\;
	\int_{0}^{\rho_0}\frac{d\rho}{\sqrt{\tfrac12(E^2-1)}}
	<\infty,
\end{equation}
so the geodesic reaches $\rho=0$ in \emph{finite} proper time.

Assuming we try to extend the \emph{same} functional form of $f$ to $\rho<0$:
\begin{equation}
	f(\rho)=1-\frac{1}{\eta\,\rho}\exp\!\left[-\frac{\eta}{2\rho^\ell}\right],\qquad \rho<0.
	\label{eq:same-form}
\end{equation}
For $\ell\notin\mathbb{Z}$ we must interpret $\rho^\ell$ on $\rho<0$ via the principal branch:
$\rho^\ell = |\rho|^\ell e^{i\pi\ell}$. Then
\begin{equation}
	-\frac{\eta}{2\rho^\ell}
	= -\frac{\eta}{2|\rho|^\ell}e^{-i\pi\ell}
	= -\frac{\eta\cos(\pi\ell)}{2|\rho|^\ell}\;+\; i\,\frac{\eta\sin(\pi\ell)}{2|\rho|^\ell},
\end{equation}
and hence
\begin{equation}
	\exp\!\left[-\frac{\eta}{2\rho^\ell}\right]
	= \exp\!\left[-\frac{\eta\cos(\pi\ell)}{2|\rho|^\ell}\right]\;
	\exp\!\left[i\,\frac{\eta\sin(\pi\ell)}{2|\rho|^\ell}\right].
\end{equation}
If $\ell\notin\mathbb{Z}$ then $\sin(\pi\ell)\neq 0$, so the second factor is a highly oscillatory \emph{complex} phase. Therefore the right-hand side of Eq. \eqref{eq:same-form} is \emph{complex-valued} for all sufficiently small negative $\rho$, implying there is no real-valued $f$ on $(-\varepsilon,0)$ that coincides with Eq. \eqref{eq:metric} on $(0,\varepsilon)$.

In conclusion we have shown that radial timelike geodesics reach $\rho=0$ in finite proper time. Subsequently, by examining the same construction for negative radial coordinate $\rho<0$  we have shown that there is no real Lorentzian extension even of class $C^0$ across $\rho=0$ that keeps the same functional form when $\ell\notin\mathbb{Z}$. Therefore, the spacetime defined by Eq. \eqref{eq:metric} is \emph{timelike geodesically incomplete} for non-integer $\ell$.

\subsubsection{Odd Integer Power}

If $\ell\in 2\mathbb{N}+1$, then the same unregulated  $f(\rho,\eta,\ell)$ for $\rho<0$ creates a \emph{hard obstruction}: near $\rho\to 0^-$ the lapse $f(\rho,\eta,\ell)\to +\infty$, making the radial timelike equation Eq. \eqref{eq:radial} inadmissible since the right-hand side is negative in any punctured neighbourhood of $0^-$. Therefore, radial timelike geodesics arriving at $\rho=0$ in finite proper time from $\rho>0$ cannot be continued to $\rho<0$; the spacetime is \emph{geodesically incomplete} at the centre.

\textbf{Proof:} For $\rho<0$ and odd $\ell$, $\rho^\ell<0$, hence the power in the exponential part of the potential in the limit approaching zero from below is 
\begin{equation}
	\lim_{\rho\to 0^-}\left(-\frac{\eta}{2\,\rho^\ell}\right)= +\infty\,,
\end{equation}
this causes the exponential term itself to satisfy 
\begin{equation}
	\lim_{\rho\to 0^-}\exp\left[-\frac{\eta}{2\,\rho^\ell}\right]= +\infty.
\end{equation}
Furthermore, the other term in the potential is also negative $1/(\eta\rho)<0$ for negative radial coordinate $\rho<0$. Therefore the limit of our potential approaching zero from the negative is 
\begin{equation}
	\lim_{\rho\to 0^-}  \frac{1}{\eta\,\rho}\exp\!\left[-\frac{\eta}{2\,\rho^\ell}\right]
	= -\infty\,.
\end{equation}
Thus, as $\rho\to 0^-$, the lapse function satisfies 
\begin{equation}
	f(\rho)=1-\frac{1}{\eta\,\rho}\exp\!\left[-\frac{\eta}{2\,\rho^\ell}\right]
	\ \xrightarrow[\rho\to 0^-]{}\ +\infty.
\end{equation}
Insert this into the radial timelike equation presented in Eq. \eqref{eq:radial}:
\begin{equation}
	\dot\rho^{\,2}=\frac{c^2}{\lambda^2}\left[E^2-f(\rho)\right]\ \xrightarrow[\rho\to 0^-]{}\ -\infty.
\end{equation}
Hence there exists $\delta>0$ such that for all $\rho\in(-\delta,0)$ the right-hand side of Eq. \eqref{eq:radial} is \emph{strictly negative}, i.e.\ there is \emph{no real} radial timelike geodesic segment in any neighbourhood of $0^-$ with the same energy $E$. Since by Eq. \eqref{eq:finite-time} a radial timelike geodesic from $\rho>0$ reaches $\rho=0$ in finite proper time, but there is no admissible continuation to $\rho<0$, the spacetime fails geodesic completeness at the centre.

\subsubsection{ Even integer power }

If $\ell\in 2\mathbb{N}$, the function $f(\rho,\eta,\ell)$ defined in Eq. \eqref{eq:metric} is $C^\infty$-extendable at $\rho=0$ by setting $f(0):=1$. Consequently, the lapse function admits a $C^{\infty}$
 extension through $\rho=0$, and the corresponding geodesic equation has locally Lipschitz coefficients across the origin $\rho=0$. Therefore, timelike geodesics that reach $\rho=0$ in finite proper time continue uniquely to $\rho<0$. Hence the spacetime is \emph{geodesically complete at the centre}.

\textbf{Proof:} For $\ell$ even and $\rho<0$, we have $\rho^\ell=|\rho|^\ell>0$, so
\begin{equation}
	h(\rho,\eta,\ell):=\exp\!\left[-\frac{\eta}{2\,\rho^\ell}\right],\qquad \rho\ne 0.
\end{equation}
It is standard that $h\in C^\infty(\mathbb{R})$ with $h^{(k)}(0)=0$ for all $k$. Since the function and all its derivatives vanish at zero we can say the function is \emph{flat} at zero. Considering the simplification
\begin{equation}
	g(\rho,\eta,\ell):=\frac{h(\rho,\eta,\ell)}{\rho},\qquad \rho\ne 0.
\end{equation}
Using flatness, for any $N$ there exists $C_N$ with $|h(\rho)|\le C_N|\rho|^{N+1}$ near $0$, hence $|g(\rho)|\le C_N|\rho|^N\to 0$.  Similar  estimates apply to derivatives of $g$, because each derivative is a linear combination of terms $h^{(k)}(\rho)/\rho^m$ that remain $O(|\rho|^M)$ for arbitrary $M$. Thus $g$ extends smoothly by $g(0,\eta,\ell):=0$. Therefore
\begin{equation}
	f(\rho,\eta, \ell)=1-\frac{1}{\eta}\,g(\rho,\eta,\ell)
\end{equation}
is $C^\infty$ at $\rho=0$ with $f(0,\eta,\ell)=1$, therefore the metric coefficients and their first derivatives are continuous at $\rho=0$. The geodesic equation itself has locally Lipschitz coefficients, so by Picard--Lindel\"of the radial timelike geodesic extends uniquely through $\rho=0$.

\subsection{Eddington--Finkelstein Diagrams}
\label{subsec:EF_diagrams}
To analyze the causal structure, we introduce a dimensionless tortoise coordinate
\begin{equation}
	\rho_{*}(\rho)=\int^{\rho}\frac{d\tilde\rho}{f(\tilde\rho)}
	\label{eq:tortoise_def}
\end{equation}
Evaluating it using the definition for $f(\rho)$ presented in Eq. \eqref{eq:g00dimless} we get the following function for the  dimensionless tortoise coordinate
\begin{equation}
\rho_*(\rho)
=
\frac{\rho_-}{
1-\dfrac{\ell\eta}{2\rho_-^\ell}
}
\ln\left|\rho-\rho_-\right|
+
\frac{\rho_+}{
1-\dfrac{\ell\eta}{2\rho_+^\ell}
}
\ln\left|\rho-\rho_+\right|
+
R(\rho)\,,
\end{equation}
which diverges logarithmically at each simple horizon. The remaining non-divergent parts $R(\rho)$ are 
\begin{equation}
R(\rho)
=
\int^\rho d\tilde{\rho}\,
\left[
\frac{1}{
1-\dfrac{1}{\eta\tilde{\rho}}
\exp\!\left(
-\dfrac{\eta}{2\tilde{\rho}^{\ell}}
\right)
}
-
\frac{1}{
f'(\rho_-)(\tilde{\rho}-\rho_-)
}
-
\frac{1}{
f'(\rho_+)(\tilde{\rho}-\rho_+)
}
\right].
\end{equation}
 Using the fact that $r_*=\lambda\rho_*$ the corresponding tortoise, light cone, and time-coordinate $t^*$ are given
by
\begin{align}
	v &\equiv t+\frac{r_*}{c}=t+\frac{\lambda}{c}\rho_*,
	\label{eq:advanced_time}\\
    u &\equiv t-\frac{r_*}{c}=t-\frac{\lambda}{c}\rho_*\\
    t^* &= v-\frac{r}{c}=t+\frac{r_*}{c}-\frac{r}{c}=t+\frac{\lambda}{c}\rho_*-\frac{r}{c}
\end{align}
the metric Eq. \eqref{eq:metric} in the ingoing coordinate $v$ and outgoing coordinate $u$, respectively, becomes
\begin{align}
	ds^{2}&=-f(\rho)c^2\,dv^{2}+2\,c\lambda dv\,d\rho+\lambda^2\rho^{2}d\Omega^{2},
	\label{eq:ef_metric_ingoing}\\
    ds^{2}&=-f(\rho)c^2\,du^{2}-2\,c\lambda du\,d\rho+\lambda^2\rho^{2}d\Omega^{2},
	\label{eq:ef_metric_outgoing}
\end{align}
which is regular at the horizons. Radial null curves ($d\Omega=0$, $ds^{2}=0$) satisfy
\begin{equation}
	0=-f(\rho)\,c^2\,dv^{2}+2c\lambda\,dv\,d\rho
	\quad\Longrightarrow\quad
	\frac{d\rho}{dv}=\frac{c}{2\lambda}f(\rho).
	\label{eq:null_rays_eq}
\end{equation}
Thus, the ingoing family is given by $dv=0$,
\begin{equation}
	v=\text{const}
	\label{eq:ingoing_null}\,.
\end{equation}
The outgoing family integrates to
\begin{equation}
	u=t^*+\frac{r}{c}-\frac{2\lambda}{c}\rho_*(\rho)=const,
	\label{eq:outgoing_null}
\end{equation}
with constants labelling distinct ingoing and outgoing rays. In the two--horizon case, the horizons $r_\pm=\lambda\rho_\pm$ appear as vertical lines in the $(t^*,r)$ plane, and the spacetime is naturally partitioned into three radial regions. The outgoing rays Eq. \eqref{eq:outgoing_null} develop the characteristic steepening as $r\to r_\pm$ associated with the logarithmic behaviour of $\rho_*$, while the ingoing rays Eq. \eqref{eq:ingoing_null} remain manifestly regular.

\begin{figure}
		\centering
		\begin{subfigure}[b]{0.49\textwidth}
			\includegraphics[width=\textwidth]{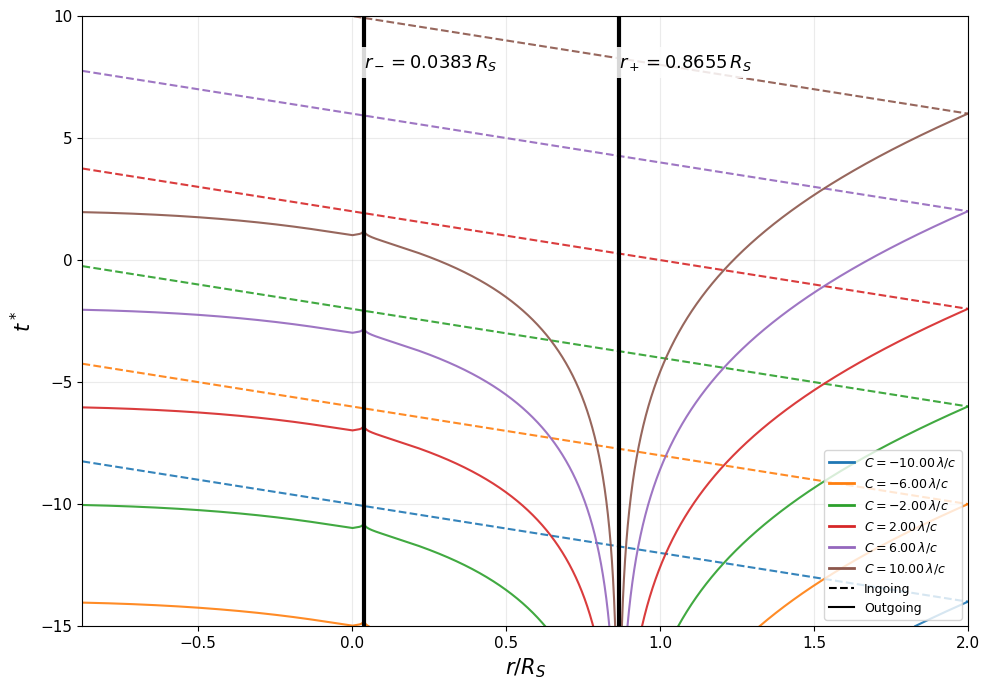}
			\caption{$\ell=1$}
			\label{fig:EFsub1}
		\end{subfigure}
		\hfill
		\begin{subfigure}[b]{0.49\textwidth}
			\includegraphics[width=\textwidth]{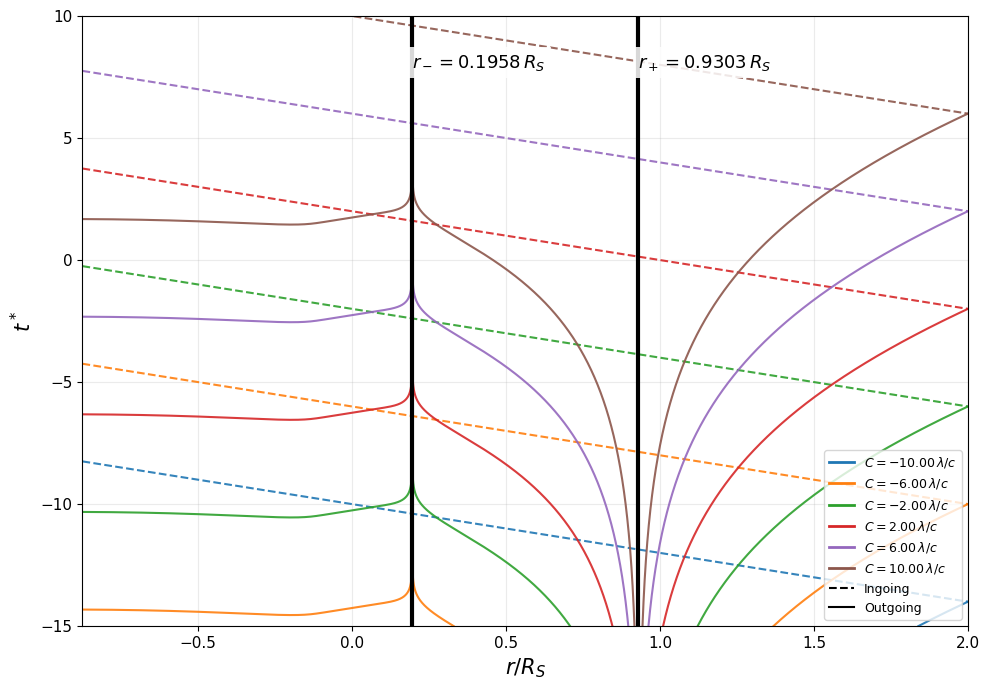}
			\caption{$\ell=2$}
			\label{fig:EFsub2}
		\end{subfigure}
	\caption{The plot shows the Eddington--Finkelstein null structure for $\eta=0.5$.
		Panel \ref{fig:EFsub1} shows $\ell=1$ with horizons at $r_-\simeq 0.0383R_s$ and $r_+\simeq 0.8655R_S$. Panel
		\ref{fig:EFsub2} shows $\ell=2$ with horizons at $r_-\simeq 0.1958R_S$ and $r_+\simeq0.9303R_S$.
		The straight dashed lines are ingoing null rays while curved lines are outgoing null rays. Each colour represents a different values of the integration constant $C$. 
		Solid vertical lines mark the horizons.}
	\label{fig:ef_diagrams_eta05}
\end{figure}

Figure \ref{fig:ef_diagrams_eta05} shows the resulting Eddington--Finkelstein diagrams for $\eta=0.5$ with $\ell=1$ and $\ell=2$. The tortoise coordinate Eq. \eqref{eq:tortoise_def} was evaluated numerically piecewise in each region, excluding a small neighbourhood of each horizon because the tortoise coordinate diverges logarithmically as $\rho\to\rho_{\pm}$

\section{Modified action}
\label{sec:fR_reconstruction_rho}

In this Section we attempt to reconstruct a gravitational action using the methodology outlined in  \cite{Das:2024tme}.
Our goal is to reconstruct, within the same weak-field perturbative scheme, a
\emph{vacuum} $f(R)$ configuration that reproduces to leading order the asymptotic behaviour outlined in Section \ref{Sec:PotentialFamily}. To simplify the calculations, we set $c=1$ only in this section.

The starting point of this calculation is the metric $f(R)$ gravity with action
\begin{equation}
	S=\frac{1}{16\pi G}\int d^4x\,\sqrt{-g}\,f(R),
	\quad
	F(R)\equiv \frac{df}{dR}.
\end{equation}
The corresponding vacuum equations are given by 
\begin{equation}
	F R_{\mu\nu}-\frac{1}{2}f g_{\mu\nu}
	-\left(\nabla_\mu\nabla_\nu-g_{\mu\nu}\Box\right)F=0.
\end{equation}
Assuming a static spherically symmetric spacetime the metric can be written in the following ansatz
\begin{equation}
	g_{\mu\nu}=\mathrm{diag}\Big(-s(r),\,p(r),\,r^2,\,r^2\sin^2\theta\Big),
\end{equation}
and we identify the weak-field potential $V(r)$ through 
\begin{equation}
	s(r)=|g_{00}|=1+2V(r)\equiv 1+\delta s(r). 
\end{equation}
The  target potential is given in Eq. \eqref{Eq:GeneralPotnetial}
\begin{equation}
	V_C(r)= -\frac{1}{2\eta\rho}\exp\!\left[-\frac{\eta}{2\rho^\ell}\right],
	\quad
	\rho=\frac{r}{\lambda},
	\quad
	\eta=\frac{\lambda}{R_s}.
\end{equation}
The metric perturbation entering the reconstruction is
\begin{equation}
	\delta s(r)=2V_C(r)= -\frac{1}{\eta\rho}\exp\!\left[-\frac{\eta}{2\rho^\ell}\right].
\end{equation}
Defining the auxiliary product
\begin{equation}
	X(r)\equiv s(r)p(r),
\end{equation}
we expand the metric in the weak-field limit $s(r)=1+\delta s(r),
X(r)=1+\delta X(r),
F(R)=1+\delta F(R).$ This yields a direct relation between the scalaron perturbation and the metric perturbation,
\begin{equation}
	\delta F(R(r))=-\delta s(r)+A+\frac{B}{r},
\end{equation}
where $A$ and $B$ are integration constants absorbing the constants in $V(r)$. The chosen potential $V_C$ satisfies the limit $V_C(r)\to 0$ as $r\to 0$, granted by the fact that 
$e^{-\eta/(2\rho^\ell)}$ suppresses $1/r$ faster than any power. Therefore, the perturbation
$\delta s(r)\to 0$ and all derivatives vanish at the origin.
To keep $F(R)$ finite at $r\to 0$, we discard the homogeneous solution
$\delta F\sim B/r$ by setting $B=0$, because it diverges at small $r$.

 The integrated relation for $\delta X$ then reads
\begin{equation}
	\delta X(r)= r\,\delta F'(r)-\delta F(r)+A,
\end{equation}
where primes denote $d/dr$.

To recover GR at low curvature or in other words $F(R)\to 1$ as $R\to 0^-$,  on the asymptotically flat branch $r\to\infty$,  we choose $A=0$ so that $\delta F\to 0$ when $V\to 0$.

With these choices out of the way, we can start working on the linear perturbation $\delta F(r)$
\begin{equation}
	\delta F(r)=-\delta s(r)=\frac{1}{\eta\rho}\exp\!\left[-\frac{\eta}{2\rho^\ell}\right],
\end{equation}
where $F(R(r))=1+\delta F(r)$. Then we
use the fact that $\delta X=r\delta F'-\delta F$ and $\delta F=-\delta s$,
\begin{equation}
	\delta X(r)=-r\,\delta s'(r)+\delta s(r),
\end{equation}
since $p=X/s$, to linear order we find
\begin{equation}
	p(r)=\frac{1+\delta X}{1+\delta s}\simeq 1+\delta p(r),
\end{equation}
remembering that $\delta p(r)=\delta X(r)-\delta s(r)=-r\,\delta s'(r)$. For the present, it is convenient to differentiate $\delta s(\rho)$ with respect to $\rho$ using $r=\lambda\rho$:
\begin{equation}
	\delta s(\rho)= -\frac{1}{\eta\rho}e^{-u},
	\quad
	u(\rho)\equiv \frac{\eta}{2\rho^\ell}.
\end{equation}
One finds
\begin{equation}
	\rho\,\frac{d\delta s}{d\rho}
	=
	e^{-u}\left(\frac{1}{\eta\rho}-\frac{\ell}{2\rho^{\ell+1}}\right),
\end{equation}
hence
\begin{equation}
	\delta p(\rho)=-\rho\,\delta s_{,\rho}
	=
	-e^{-u}\left(\frac{1}{\eta\rho}-\frac{\ell}{2\rho^{\ell+1}}\right).
\end{equation}

We write the diagonal metric ansatz using the linearized weak-field expansion
\begin{equation}
	g_{\mu\nu}=\mathrm{diag}\big(-(1+\delta s),\,1+\delta p,\,r^2,\,r^2\sin^2\theta\big),
\end{equation}
from which we find the Ricci scalar expanded to first order in the linear perturbations $\delta s,\delta p$ 
\begin{equation}
	R(r)= -\delta s''(r)+\frac{2}{r}\delta p'(r)-\frac{2}{r}\delta s'(r)+\frac{2}{r^2}\delta p(r),
\end{equation}
and with $\delta p=-r\,\delta s'$ this simplifies to
\begin{equation}
	R(r)=-3\left(\delta s''(r)+\frac{2}{r}\delta s'(r)\right).
\end{equation}

For the potential chosen in Eq. \eqref{Eq:GeneralPotnetial}
\begin{equation}
	\delta s(\rho)= -\frac{1}{\eta\rho}e^{-u},
	\quad
	u=\frac{\eta}{2\rho^\ell},
\end{equation}
direct differentiation yields the closed form
\begin{equation}\label{eq:RicciScalrClosedForm}
	R(r)=\frac{3\ell}{4\lambda^2}\,\rho^{-2\ell-3}
	\left(\ell\eta-2(\ell+1)\rho^\ell\right)
	\exp\!\left[-\frac{\eta}{2\rho^\ell}\right],
\end{equation}
where $\rho=\frac{r}{\lambda}$

Then we can easily verify that the Ricci scalar approaches zero at both the origin and large distances, thus satisfying the asymptotic conditions imposed in Sec. \ref{Sec:PotentialFamily}
\begin{equation}
	\lim_{\rho\rightarrow 0^+}R(\rho) = 0^+ \quad \text{and} \quad
	\lim_{\rho\rightarrow\infty}R(\rho) = 0^-.
\end{equation}
An interesting observation is that the Ricci scalar $R$ changes sign at
\begin{equation}
	\rho^\ell=\frac{\ell\eta}{2(\ell+1)},
	\qquad\text{equivalently}\qquad
	u=\frac{\ell+1}{\ell}.
\end{equation}

In to find the $F(R)$ function we need to obtain the radial coordinate as a function of the Ricci scalar. This is done by finding the inverse function of Eq. \eqref{eq:RicciScalrClosedForm}. 

First we introduce a dimensionless variable
\begin{equation}
	u\equiv \frac{\eta}{2\rho^\ell}
	=\frac{\eta}{2}\left(\frac{\lambda}{r}\right)^\ell,
	\quad
	\rho=\left(\frac{\eta}{2u}\right)^{1/\ell},
	\quad
	r=\lambda\left(\frac{\eta}{2u}\right)^{1/\ell}.
\end{equation}
In $u$-form, the curvature and the scalaron read
\begin{align}
	\lambda^2 R(u)&=3\,2^{3/\ell}\,\ell\,\eta^{-(1+3/\ell)}
	u^{1+3/\ell}\big(\ell u-\ell-1\big)e^{-u},\\
	F(u)&=1+\delta F(u)=1+\eta^{-(1+1/\ell)}(2u)^{1/\ell}e^{-u}\,.
\end{align}
These equations provide parametric expressions for $R$ and $F$ along the radius $r$.

In general the equation
\begin{equation}
	\lambda^2 R=C\,u^{1+3/\ell}(\ell u-\ell-1)e^{-u}
\end{equation}
for $u=u(R)$ is transcendental, so exact inversion requires either numerical root-finding or special-function generalized Lambert inverses.
However, a controlled analytic inversion is available in a few different regimes.

For example, on the weak-curvature branch, corresponding to large radius $r$, small variable $u\ll 1$, and negative Ricci scalar $R<0$, we may use the approximation $e^{-u}\simeq 1$ and $\ell u-\ell-1\simeq-(\ell+1)$. The Ricci scalar then becomes
\begin{equation}
	R(r)\simeq -\frac{3\ell(\ell+1)}{2}\frac{\lambda^{\ell+1}}{r^{\ell+3}}
\end{equation}
leading to the following for the radial coordinate
\begin{equation}
	r(R)\simeq
	\left[
	\frac{3\ell(\ell+1)}{2}\frac{\lambda^{\ell+1}}{-R}
	\right]^{1/(\ell+3)}.
\end{equation}
Equivalently,
\begin{equation}
	u(R)\simeq
	\frac{1}{2}\left(\frac{2\lambda^2}{3\ell(\ell+1)}\right)^{\ell/(\ell+3)}
	\eta\,(-R)^{\ell/(\ell+3)}\,.
\end{equation}
At the origin, in other words small radius $r$, large variable $u\gg 1$, and a positive Ricci scalar $R\to 0^+$), we find that the Ricci scalar can be written as 
\begin{equation}\label{eq:RicciSmallr}
	R(u)\sim u^{2+3/\ell}e^{-u}\,.
\end{equation}
As evident from the above, the Ricci scalar exponentially approaches zero.
The inversion of the function in Eq. \eqref{eq:RicciSmallr} is a logarithm to leading order
\begin{equation}
	u \sim \ln\!\left(\frac{R_\ast}{R}\right)
	+\left(2+\frac{3}{\ell}\right)\ln u+\cdots,
\end{equation}
with $R_\ast$ a scale set by $\lambda,\eta,\ell$.

The components needed to integrate to obtain $f(R)$ are all present. Following the formula
\begin{equation}
	\delta f(r)=\int dR(r)\,\delta F(R(r))
	=\int dr\,\delta F(r)\,\frac{dR}{dr}.
\end{equation}
The parametrization in terms of the variable $u$ gives $dR=(dR/du)\,du$, so
\begin{equation}
	\delta f(u)=\int \delta F(u)\,\frac{dR(u)}{du}\,du.
\end{equation}

This integral can be done in closed form for all $\ell>0$ because
\begin{equation}
	\delta F(u)\propto u^{1/\ell}e^{-u},
	\quad
	R(u)\propto u^{1+3/\ell}(\ell u-\ell-1)e^{-u},
\end{equation}
making the integrand a polynomial in $u$ times $u^{4/\ell}e^{-2u}$.
The result is naturally expressed with the upper incomplete gamma function $\Gamma(\alpha,u)$.

One first defines the coefficients
\begin{subequations}
	\begin{align}
		\alpha\equiv \frac{4}{\ell},
		&\qquad
		a_2=-\ell^2,
		\\
		a_1=3\ell^2+4\ell,
		&\qquad
		a_0=-(\ell+1)(\ell+3).
	\end{align}
\end{subequations}
Then one convenient form, fixing the integration constant so that
$\delta f\to 0$ as $u\to 0$ and hence $f(R)\to R$ as $R\to 0$, is
\begin{equation}
	f(R(u))=R(u)+\delta f(u),
\end{equation}
\begin{align}
		\delta f(u)
		=-&\frac{3\,\eta^{-(2+4/\ell)}}{\lambda^2}
		\left[
		\frac{a_2}{8}\left(\Gamma(\alpha+3,2u)-\Gamma(\alpha+3)\right)
		\right.\nonumber\\
		&+\left.\frac{a_1}{4}\left(\Gamma(\alpha+2,2u)-\Gamma(\alpha+2)\right)+\frac{a_0}{2}\left(\Gamma(\alpha+1,2u)-\Gamma(\alpha+1)\right)
		\right].
\end{align}
One arrives at a parametric reconstruction of the function $f(R)$ within the present linearized weak-field scheme. Hence the reconstruction is represented parametrically by the curve $u\mapsto (R(u),f(u))$.

\subsection{Approximate action at large distances}

In the weak-field limit $u\ll 1$, or equivalently large  distances $r$ compared to the additional scale $\lambda$, the Ricci scalar and the inversion $u$ are perturbative.
Using the leading inversion above we obtain
\begin{equation}
		F(R)=1+\frac{1}{\eta}
		\left(\frac{2\lambda^2}{3\ell(\ell+1)}\right)^{1/(\ell+3)}
		(-R)^{1/(\ell+3)}+O\!\left(|R|^{2/(\ell+3)}\right)
\end{equation}
and
\begin{equation}
		f(R)=R-\frac{\ell+3}{\ell+4}\,\frac{1}{\eta}
		\left(\frac{2\lambda^2}{3\ell(\ell+1)}\right)^{1/(\ell+3)}
		(-R)^{(\ell+4)/(\ell+3)}+O\!\left(|R|^{(\ell+5)/(\ell+3)}\right)
\end{equation}

These asymptotics show explicitly that $f(R)\to R$ as $|R|\to 0$, with a non-analytic fractional-power correction controlled by $\ell$ and suppressed by the ratio $\eta=\lambda/R_s$.
%
%
%

Since $R(u)$ changes sign at $u=(\ell+1)/\ell$, the inversion $u(R)$ and  $f(R)$ are not globally single-valued functions, and thus require branch restrictions.
The weak-curvature series above corresponds to the $R<0$ branch near $u\to 0$.
The additive constant $A$ in $f$ is set so that the limit of $f(R)\to R$ as $R\to 0$, corresponding to no induced cosmological constant in the low-curvature limit.

\subsection{Gravitational action close to the origin}

For the chosen potential in Eq. \eqref{Eq:GeneralPotnetial},
the short-distance regime corresponds to
\begin{equation}
	\rho\to 0^+,
	\quad\Longleftrightarrow\quad
	u\equiv \frac{\eta}{2\rho^\ell}\to\infty .
\end{equation}
From
\begin{align}
	\delta s(r)&=2V_C(r)= -\frac{1}{\eta\rho}\,e^{-u},\\
	\delta F(r)&=-\delta s(r)=\frac{1}{\eta\rho}\,e^{-u},
\end{align}
and from the linearized Ricci scalar derived in Eq. \eqref{eq:RicciScalrClosedForm},
one obtains the Ricci scalar at leading order,
\begin{equation}
	R(\rho)\sim \frac{3\ell^2\eta}{4\lambda^2}\,\rho^{-2\ell-3}e^{-u},
	\quad
	\delta F(\rho)\sim \frac{1}{\eta\rho}e^{-u}.
\end{equation}

It is convenient to rewrite these in terms of $u=\eta/(2\rho^\ell)$. Since
\begin{equation}
	\rho=\left(\frac{\eta}{2u}\right)^{1/\ell},
\end{equation}
we find 
\begin{equation}
	R(u)\sim B\,u^\beta e^{-u},
	\qquad
	\delta F(u)\sim D\,u^{1/\ell}e^{-u},
\end{equation}
where
\begin{equation}
	\beta=2+\frac{3}{\ell},
	\quad
	B=\frac{3\,2^{3/\ell}\ell^2}{\lambda^2}\,\eta^{-(1+3/\ell)},
	\quad
	D=2^{1/\ell}\eta^{-(1+1/\ell)}.
\end{equation}
Hence
\begin{equation}
	\delta F(R)\sim \frac{D}{B}\,R\,u^{-(2+2/\ell)},
\end{equation}
namely
\begin{equation}
	\delta F(R)\sim
	\frac{\lambda^2\eta^{2/\ell}}{3\,2^{2/\ell}\ell^2}\,
	R\,u^{-(2+2/\ell)}.
\end{equation}
To express this in terms of $R$, observe that the relation
\begin{equation}
	R\sim B\,u^\beta e^{-u}
\end{equation}
which implies that
\begin{equation}
	u-\beta\ln u=\ln\!\left(\frac{B}{R}\right).
\end{equation}
Therefore, it is useful to define
\begin{equation}
	L(R)\equiv \ln\!\left(\frac{B}{R}\right)
	=
	\ln\!\left(
	\frac{3\,2^{3/\ell}\ell^2}{\eta^{1+3/\ell}\lambda^2R}
	\right),
\end{equation}
we arrive at the following expression for the asymptotic inversion
\begin{equation}
	u(R)=L+\beta\ln L+O\!\left(\frac{\ln L}{L}\right),
\end{equation}
Substituting this into $\delta F$ yields the core-branch scalaron
\begin{equation}
	F(R)=1+\delta F(R),
\end{equation}
with
\begin{equation}
	F(R)\sim
	1+
	\frac{\lambda^2\eta^{2/\ell}}{3\,2^{2/\ell}\ell^2}\,
	\frac{R}{\big[L(R)\big]^{2+2/\ell}}
	\left[
	1+O\!\left(\frac{\ln L}{L}\right)
	\right]\,,
\end{equation}

Since $F(R)=df/dR$, integration gives the short-distance action.
Choosing the additive constant so that $f(0)=0$ on this branch, we find
\begin{equation}
	f(R)\sim
	R+
	\frac{\lambda^2\eta^{2/\ell}}{6\,2^{2/\ell}\ell^2}\,
	\frac{R^2}{\big[L(R)\big]^{2+2/\ell}}
	\left[
	1+O\!\left(\frac{\ln L}{L}\right)
	\right]\,,
\end{equation}
where
\begin{equation}
	L(R)=
	\ln\!\left(
	\frac{3\,2^{3/\ell}\ell^2}{\eta^{1+3/\ell}\lambda^2R}
	\right).
\end{equation}
It is interesting that for the gravitational action close to the core the correction is not a fractional power of $R$, but a
logarithmically screened $R^2$ term.

\subsection{Action for $\ell=2$}
In the case $\ell=2$ is the lowest even integer power for which  the potential yields a resolution of the singularity and a geodesically complete spacetime. This particular case is discussed here.

\textbf{In the weak-field limit for $\ell=2$},
\begin{equation}
	u=\frac{\eta}{2\rho^2}=\frac{\eta\lambda^2}{2r^2}.
\end{equation}
We find the reconstructed components of the metric
\begin{equation}
	\delta s(\rho)=-\frac{1}{\eta\rho}e^{-u},
	\qquad
	F(u)=1+\eta^{-3/2}\sqrt{2u}\,e^{-u}.
\end{equation}
The Ricci scalar simplifies to 
\begin{equation}
	R(\rho)=\frac{3}{\lambda^2}\,\rho^{-7}\big(\eta-3\rho^2\big)e^{-u},
	\quad
	u=\frac{\eta}{2\rho^2}.
\end{equation}
The sign change where $R=0$ occurs at
\begin{equation}
	\rho=\sqrt{\frac{\eta}{3}},
	\qquad\text{i.e.}\qquad
	u=\frac{3}{2}.
\end{equation}

In the $\ell=2$ case, the incomplete gamma orders
are integers, and thus give a closed expression for the function $\delta f(u)$, and reduces to an elementary polynomial multiplied by $e^{-2u}$ form.
Fixing $\delta f(0)=0$,
\begin{equation}
	\delta f(u)=\frac{3}{\eta^4\lambda^2}
	\left[
	3+\big(8u^4-24u^3-6u^2-6u-3\big)e^{-2u}
	\right],
\end{equation}
and
\begin{equation}
	f(R(u))=R(u)+\delta f(u)
\end{equation}
parametrically.

The weak-curvature inversion and the corresponding series for  $F(R)$, and $f(R)$ on the $u\ll 1$ branch are given by the relation,
\begin{equation}
	R(r)\simeq -9\frac{\lambda^3}{r^5}
	\quad \Longrightarrow \quad
	r(R)\simeq \left(\frac{9\lambda^3}{-R}\right)^{1/5}
	.
\end{equation}
Then the $f(R)$ and $F(R)$ functions defining the weak-field action are 
\begin{equation}
	F_{\rho\rightarrow\infty}(R)=1+\frac{1}{\eta}\left(\frac{\lambda^2}{9}\right)^{1/5}(-R)^{1/5}
	+O(|R|^{2/5})
\end{equation}
and
\begin{equation}\label{eq:l2_fR_series_R2}
	f_{\rho\rightarrow\infty}(R)=R-\frac{5}{6}\,\frac{1}{\eta}\left(\frac{\lambda^2}{9}\right)^{1/5}(-R)^{6/5}
	+O(|R|^{7/5})\,.
\end{equation}

\textbf{Near the origin for $\ell=2$}, the Ricci scalar on the same branch is
\begin{equation}
	R(\rho)=\frac{3}{\lambda^2}\rho^{-7}\big(\eta-3\rho^2\big)e^{-u}.
\end{equation}
Hence, as $\rho\to0^+$ (equivalently $u\to\infty$),
\begin{equation}
	R(u)\sim \frac{24\sqrt{2}}{\lambda^2\eta^{5/2}}\,u^{7/2}e^{-u}.
\end{equation}
Therefore
\begin{equation}
	\delta F(R)\sim \frac{\lambda^2\eta}{24}\,\frac{R}{u^3}.
\end{equation}

To invert $R(u)$, define
\begin{equation}
	B\equiv \frac{24\sqrt{2}}{\lambda^2\eta^{5/2}},
	\qquad
	L(R)\equiv \ln\!\left(\frac{B}{R}\right)
	=
	\ln\!\left(\frac{24\sqrt{2}}{\eta^{5/2}\lambda^2R}\right).
\end{equation}
Since
\begin{equation}
	R\sim B\,u^{7/2}e^{-u},
\end{equation}
we find that 
\begin{equation}
	u-\frac{7}{2}\ln u = L(R),
\end{equation}
and therefore the asymptotic inversion
\begin{equation}\label{eq:loginvestion}
	u(R)=L+\frac{7}{2}\ln L + O\!\left(\frac{\ln L}{L}\right),
	\qquad R\to0^+.
\end{equation}
Substituting this into $\delta F$ yields the short-distance scalaron
\begin{equation}
	F(R)\sim
	1+\frac{\lambda^2\eta}{24}\,
	\frac{R}{[L(R)]^3}
	\left[1+O\!\left(\frac{\ln L}{L}\right)\right]
	.
\end{equation}

Integrating with respect to $R$, and choosing the additive constant so that
$f(0)=0$, gives the asymptotic core action
\begin{equation}
	f_{\rho\rightarrow 0}(R)\sim
	R+\frac{\lambda^2\eta}{48}\,
	\frac{R^2}{[L(R)]^3}
	\left[1+O\!\left(\frac{\ln L}{L}\right)\right]\,,
\end{equation}
where
\begin{equation}
	L(R)=\ln\!\left(\frac{24\sqrt{2}}{\eta^{5/2}\lambda^2R}\right).
\end{equation}

Equivalently, using $\eta=\lambda/R_s$, this can be written as
\begin{equation}\label{eq:f_smallRplus22}
	f_{\rho\rightarrow 0}(R)\sim
		R+\frac{\lambda^3}{48R_s}\,
	\frac{R^2}{\left[\ln\!\left(\dfrac{24\sqrt{2}\,R_s^{5/2}}{\lambda^{9/2}R}\right)\right]^3}
	\left[1+O\!\left(\frac{\ln\ln(B/R)}{\ln(B/R)}\right)\right].
\end{equation}
This makes explicit that the same model yields \emph{two distinct small-curvature
	branches}: one at large $\rho$,
and one near the centre, where the corrections are logarithmically suppressed.

For $\ell=2$, $f(R)$ cannot be represented globally as a single function, because that $R(\rho)$ is not globally invertible. Therefore, the reconstruction needs to be approached branch by branch. On the asymptotically flat weak-curvature branch $R\to 0^-$ we find
the explicit series Eq. \eqref{eq:l2_fR_series_R2}, while near the centre $R\to 0^+$ one
obtains a logarithmic inversion Eq. \eqref{eq:loginvestion} and the corresponding
suppressed correction Eq. \eqref{eq:f_smallRplus22}.

\section{Conclusion}
\label{sec:Conclusion}

In this work we studied a broad class of  gravitational potentials giving rise to singularity-free theories of gravity.  We parametrized the different branches of the potential family and identified the branch
which reproduces the Schwarzschild behaviour at large distances and softens the short-distance gravitational interaction.  We embedded the potentials into a weak-field approximation of the metric, further showing  that this leads to a spacetime that is asymptotically flat, has vanishing force and curvature in the appropriate limits, and provides a concrete setting in which the distinction between curvature regularity and true singularity resolution can be tested explicitly. 

Our analysis shows that these geometries generically possess a two-horizon structure. The horizon radii are  written in closed form in terms of the Lambert $W$ function, and we showed that the existence of horizons is controlled by the bound  $\frac{\ell}{2}\,\left(\lambda/R_S\right)^{\,\ell+1}\leq e^{-1}$, where the bound depends on both the Schwarzschild radius $R_S$ and the additional short-distance length scale $\lambda$. This in turn implies an extremal configuration and a corresponding minimum black-hole mass,
set by the additional short-distance scale $\lambda$. Therefore, in this framework the regularization scale is not merely a mathematical regulator, but has direct physical consequences for the allowed black-hole spectrum.

However, the regularity of the potential and curvature is not by itself sufficient to support the claim of resolution of the singularity. Geodesic completeness imposes a further, nontrivial restriction on the admissible potentials. Radial timelike geodesics with $E>1$ reach $\rho=0$ in finite proper time, so the decisive question is whether the geometry admits a real and unambiguous extension through the origin. We showed that for non-integer $\ell$ the continuation to $\rho<0$ becomes complex-valued and therefore fails to define a real Lorentzian geometry, while for odd integer $\ell$ the lapse diverges in such a way that no admissible continuation of radial timelike geodesics exists. By contrast, for even integer $\ell$ the metric extends smoothly through $\rho=0$, the geodesic equations locally obey the Lipschitz continuity, and radial timelike geodesics continue uniquely across the origin. In this sense, the geodesically complete sector of the theory is selected dynamically and mathematically by the requirement of a smooth extension, rather than by curvature considerations alone. 

The causal structure obtained from the tortoise coordinate and the advanced Eddington--Finkelstein construction is consistent with this picture. In the non-extremal regime the spacetime separates into the expected radial regions bounded by the inner and outer horizons, and the null-ray structure provides a transparent geometric diagnostic of the extension properties discussed analytically. This makes clear that the present framework, in addition to regularizing the invariants offers a controlled spacetime model in which horizon structure, null ray propagation, and geodesic evolution can all be analyzed in a unified way. 

Finally, we showed that the weak-field potential can be used to reconstruct a corresponding $f(R)$ theory of gravity whose vacuum solution obeys the required asymptotics.  In general the reconstruction is naturally parametric, and for the geodesically complete case $\ell=2$, it becomes explicit that the resulting action is branch-wise rather than globally single-valued, due to the non-monotonic behaviour of $R(\rho)$. On the asymptotically flat branch the correction to Einstein--Hilbert gravity is non-analytic, while near the centre it is strongly suppressed. This provides a concrete link between the effective exponentially softened potential and a covariant modified-gravity description. 

The results of this work provide a class of gravitational potentials which resolve the singularity. Furthermore, the class of models provided by the chosen potentials have geophysically complete spacetime solutions.  The class of models studied here supplies an analytically tractable example  in which exponentially softened potentials produce regular and asymptotically flat black-hole geometries. Genuinely geodesically complete spacetimes are produced only for a restricted subclass of potentials.  A natural next step is to go beyond the present leading-order reconstruction and determine whether the same conclusions persist in a fully non-perturbative covariant theory, as well as to study stability and thermodynamic properties. An interesting further phenomenological study is to determine whether the minimum-mass configurations implied by Eq. \eqref{eq:Mmin} persist once evaporation, back-reaction, and stability are included, and whether they can then be interpreted as long-lived end states of the dynamics.
\section*{Acknowledgements}
I want to express my most sincere gratitude to Dr. Saurya Das and Dr. Mitja Fridman  for the helpful discussions, suggestions, and support which led to the crystallization of the idea into this paper. 

\begin{appendices}

\section{Reduction to the radial sector at the centre}
\label{app:radial-sector-center}

In this appendix, we show that, for the class of static spherically symmetric metrics considered here, radial geodesics are sufficient to test whether the centre $\rho=0$ is a source of geodesic incompleteness. 

For the present argument it is convenient to set $c=\lambda=1$. The metric then reads
\begin{equation}
ds^2=-f(\rho)\,dt^2+f(\rho)^{-1}d\rho^2+\rho^2 d\Omega^2 ,
\label{eq:center-test-metric}
\end{equation}
with
\begin{equation}
f(\rho)=1-\frac{1}{\eta\rho}\exp\!\left(-\frac{\eta}{2\rho^\ell}\right).
\label{eq:center-test-lapse}
\end{equation}
By spherical symmetry, every geodesic is contained in a fixed two-plane, so without loss of generality we may restrict to the equatorial plane $\theta=\pi/2$. Along any affinely parametrized geodesic there are two conserved quantities,
\begin{equation}
E=f(\rho)\,\dot t,
\qquad
L=\rho^2 \dot\phi ,
\label{eq:center-test-conserved}
\end{equation}
where dot denotes differentiation with respect to an affine parameter $s$. Writing the normalization as
\begin{equation}
g_{\mu\nu}\dot x^\mu \dot x^\nu=-\varepsilon,
\qquad
\varepsilon=
\begin{cases}
1, & \text{timelike geodesics},\\
0, & \text{null geodesics},\\
-1, & \text{spacelike geodesics},
\end{cases}
\label{eq:center-test-normalization}
\end{equation}
one obtains
\begin{equation}
-\varepsilon
=
-f(\rho)\dot t^2+f(\rho)^{-1}\dot\rho^2+\rho^2\dot\phi^2 .
\label{eq:center-test-normexpanded}
\end{equation}
Substituting Eq.~\eqref{eq:center-test-conserved} into Eq.~\eqref{eq:center-test-normexpanded} gives the effective radial equation
\begin{equation}
\dot\rho^2
=
E^2-f(\rho)\left(\varepsilon+\frac{L^2}{\rho^2}\right).
\label{eq:center-test-effective}
\end{equation}

The key observation is that for the even-$\ell$ branch discussed below, the lapse function extends smoothly to the centre and satisfies
\begin{equation}
f(0)=1.
\label{eq:center-test-f0}
\end{equation}
Hence there exist constants $\delta>0$ and $m>0$ such that
\begin{equation}
f(\rho)\ge m>0,
\qquad
|\rho|<\delta .
\label{eq:center-test-lowerbound}
\end{equation}
Assume now that a geodesic reaches $\rho=0$. If $L\neq 0$, Eq.~\eqref{eq:center-test-effective} implies
\begin{equation}
\dot\rho^2
\le
E^2+ \sup_{|\rho|<\delta} f(\rho)\,|\varepsilon|
-
m\,\frac{L^2}{\rho^2}.
\label{eq:center-test-contradiction}
\end{equation}
As $\rho\to 0$, the last term diverges to $-\infty$, so the right-hand side becomes negative for sufficiently small $|\rho|$. This is impossible for a real geodesic, since $\dot\rho^2\ge 0$. Therefore any geodesic that can reach the centre must satisfy
\begin{equation}
L=0 .
\label{eq:center-test-radial}
\end{equation}
In other words, only radial geodesics can probe the centre.

This proves that the question of whether $\rho=0$ is a source of geodesic incompleteness reduces to the radial sector. Equivalently, if all radial geodesics that reach the centre extend across $\rho=0$, then the centre does not obstruct the extension of any geodesic. Conversely, a single incomplete radial geodesic is sufficient to establish incompleteness at the centre.




\end{appendices}


\bibliography{Ref}

\end{document}